\documentclass{article}
\usepackage{spconf,amsmath,graphicx}

\usepackage{diagbox}
\usepackage{multirow}
\usepackage{multicol}
\usepackage[fontsize=9pt]{fontsize}
\usepackage{amsmath}
\usepackage{hhline}
\usepackage{arydshln}
\usepackage{enumitem}
\newlength{\bibitemsep}
\newlength{\bibparskip}
\let\oldthebibliography\thebibliography
\renewcommand\thebibliography[1]{%
  \oldthebibliography{#1}%
  \setlength{\parskip}{\bibitemsep}%
  \setlength{\itemsep}{\bibparskip}%
}
\title{Fixed-point quantization aware training for on-device keyword-spotting}
\name{%
\begin{tabular}{@{}c@{}}
Sashank Macha$^{\star}$ \qquad 
Om Oza$^{\star}$ \qquad 
Alex Escott$^{\star}$ \qquad
Francesco Caliv\'a$^{\star}$ \qquad \\
Robbie Armitano$^{\star}$ \qquad 
Santosh Kumar Cheekatmalla$^{\star}$ \qquad \\
Sree Hari Krishnan Parthasarathi$^{\dagger}$ \quad
Yuzong Liu$^{\star}$ \qquad 
\end{tabular}}
 \address{$^{\star}$Alexa Perceptual Technologies, $^{\dagger}$Alexa AI, Amazon}
\begin{document}
%
\maketitle
\begin{abstract}
Fixed-point (FXP) inference has proven suitable for embedded devices with limited computational resources, and yet model training is continually performed in floating-point (FLP). FXP training has not been fully explored and the non-trivial conversion from FLP to FXP presents unavoidable performance drop. We propose a novel method to train and obtain FXP convolutional keyword-spotting (KWS) models. We combine our methodology with two quantization-aware-training (QAT) techniques -- squashed weight distribution and absolute cosine regularization for model parameters, and propose techniques for extending QAT over transient variables, otherwise neglected by previous paradigms. Experimental results on the Google Speech Commands v2 dataset show that we can reduce model precision up to 4-bit with no loss in accuracy. Furthermore, on an in-house KWS dataset, we show that our 8-bit FXP-QAT models have a 4-6\% improvement in relative false discovery rate at fixed false reject rate compared to full precision FLP models. During inference we argue that FXP-QAT eliminates q-format normalization and enables the use of low-bit accumulators while maximizing SIMD throughput to reduce user-perceived latency. We demonstrate that we can reduce execution time by 68\% without compromising KWS model’s predictive performance or requiring model architectural changes. Our work provides novel findings that aid future research in this area and enable accurate and efficient models.
\end{abstract}
\begin{keywords}
keyword-spotting, quantization-aware-training, fixed-point-arithmetic, on-device
\end{keywords}
\vspace{-3mm}
\section{Introduction}
\label{sec:intro}
\vspace{-2mm}
We are interested in a small memory footprint and low latency keyword spotting (KWS) system for embedded on-the-go (OTG) devices. Modern KWS algorithms use deep neural networks (DNN) and have shown to achieve human-level performance~\cite{goodfellow2016deep}. Despite their success, DNN deployment on-device is challenging due to computational and bandwidth demand. These challenges can be greatly mitigated by converting expensive floating-point (FLP) operations to fixed-point (FXP) while simultaneously improving inference speed~\cite{Kouris2018CascadeCNNPT}. FXP multiplications consume 18.5$\times$ less energy and half the memory compared to FLP~\cite{6757323}. However, DNN training is usually conducted in FLP due to operation parallelism on commercial GPUs, and wide numerical range that allows for better gradient updates. While similar parallelization for FXP can be realized on FPGAs~\cite{9184436}, the lack of a processor comparable to GPUs makes training significantly slow. Most approaches thereby resort to converting models trained in FLP to FXP introducing bias that degrades performance~\cite{10.1145/3020078.3021741, 8803490}. We propose a novel method that addresses the limitations of FXP training and uses FLP workflows to obtain FXP models.\\
Computational requirements can be reduced further using low-precision inference via quantization, which allows increased operations per accessed memory byte~\cite{10.1145/3020078.3021741, Umuroglu2017FINNAF}. Such quantization is typically achieved by means of post-training-quantization (PTQ)~\cite{ptq}, which however causes severe information loss affecting model accuracy. To maintain overall accuracy for quantized DNNs, quantization can be incorporated in the training phase leading to quantization-aware-training (QAT). QAT introduces quantization noise during training by means of deterministic rounding~\cite{Mishchenko2019LowBitQA, Jacob2018QuantizationAT, Li2017TrainingQN}, reparametrization~\cite{strom2022squashed, Dupont2021WeightRF} or regularization~\cite{Nguyen2020QuantizationAT, Ding2019RegularizingAD} among few techniques, allowing DNNs to adapt to inference quantization. Notable work has shown that with QAT model parameters can be learned at binary and ternary precision~\cite{courbariaux2015binaryconnect, Zhang2020TernaryBERTDU}.\\
In this paper, we discuss methods through which we enable FXP training and combine it with QAT. Our approach allows training up-to ultra-low-precision (4-bit) FXP models that are accurate, and perform on-par with full-precision (32-bit) FLP models demonstrated on publicly available Google Speech Commands v2 dataset~\cite{warden2018speech}. On an inhouse KWS dataset, our 8-bit FXP-QAT models outperform the 32-bit FLP baseline by a relative 4-6\%. We additionally propose key inference optimizations that make KWS on constrained OTG devices possible. Our proposed approach achieves 68\% faster inference while maintaining similar accuracy to full-precision FLP models. Furthermore, we argue that with QAT, we can eliminate $\sim$10\% CPU cycles otherwise spent in normalizing activation precision. Our contributions can be summarized as follows -
\begin{itemize}[noitemsep,topsep=0pt]
    \item we propose novel FXP model training that leverages workflows optimized for FLP;
    \item we extend two QAT techniques to FXP and train low-bit KWS models with minimal accuracy degradation;
    \item we reduce inference execution time using low-bit accumulators enabled through FXP-QAT;
    \item we propose a two-tier accumulator-buffer approach to reduce overflow/saturation typical of large DNN layers; 
\end{itemize}
The remainder of the paper is organized as follows: in sections~\ref{sec:related work} and \ref{sec:qat}, we present related work, and introduce QAT focusing on two methods based on squashed weight distribution and absolute-cosine regularization. In section~\ref{sec:fixed-point inference}, we propose and describe our contributions to these methods, which allow us to achieve fixed-point training. We describe our experimental setup, and results in section~\ref{sec:exp-setup}. After discussion, we conclude the paper with our vision on the future of FXP-QAT for small footprint devices.
\vspace{-2.5mm}
\section{Related Work}
\label{sec:related work}
\vspace{-1mm}
Among inference setups aimed at enabling DNN deployment for low-power platforms, FXP systems play a key role~\cite{8451268}. FXP offered the most promising trade-off between accuracy and computational complexity among different number systems~\cite{10.1145/2925426.2926294}. Horowitz \textit{et al.}~\cite{6757323} found corroborative evidence with FXP multiply-accumulate (MAC) units compared to FLP on 45\textit{nm} technology. Notable work has been done in FXP inference — Umuroglu \textit{et al.}~\cite{Umuroglu2017FINNAF} built a framework that mapped binary weights to hardware, while other foundational works proposed to design schemes for FXP inference~\cite{10.1145/2684746.2689060, Zhao2017AcceleratingBC}. On another front, there has been work on efficient model topologies~\cite{9184436}, better approaches to quantize parameters to FXP with low errors~\cite{8803490} and methods to address limitations of FXP training~\cite{8615717, 9311563, 9383824}. While the majority of the work has been centered around developing suitable models~\cite{7424338} or efficient hardware, QAT improves model adaptation to FXP. Lin \textit{et al.}~\cite{Lin2016FixedPQ} used SQNR based quantization to achieve better accuracy with $8$-bit FXP but suffered from data overflow during inference. Anwar \textit{et al.}~\cite{7178146} presented an exhaustive L2-error minimization based quantization but required pre-training with full-precision. In the context of speech processing, $4$ to $6$-bit QAT have been studied for event detection~\cite{Shi2019CompressionOA}, speech recognition~\cite{Nguyen2020QuantizationAT} and KWS~\cite{Mishchenko2019LowBitQA} but limited to floating-point. All the above methods either require pretraining of model parameters or have runtime limitations and have limited discussion on inference computations. 
\vspace{-2mm}
\section{QUANTIZATION AWARE TRAINING}
\label{sec:qat}
\vspace{-1mm}
Quantized models are typically obtained by full-precision training followed by quantization. Nevertheless, for aggressive compression, low-bit quantization or cross format conversion from FLP to FXP, such quantization degrades accuracy. With QAT, we introduce inference specific quantization-noise during training to avoid inference time errors. Our FXP training is agnostic to QAT methodology and can extend to different technqiues. We choose two recent works in this domain and describe how we enable FXP-QAT for different model components.
\begin{figure}[tb]
  \centering
  \includegraphics[trim={1.5mm 0 2mm 0},clip,width=0.9\columnwidth]{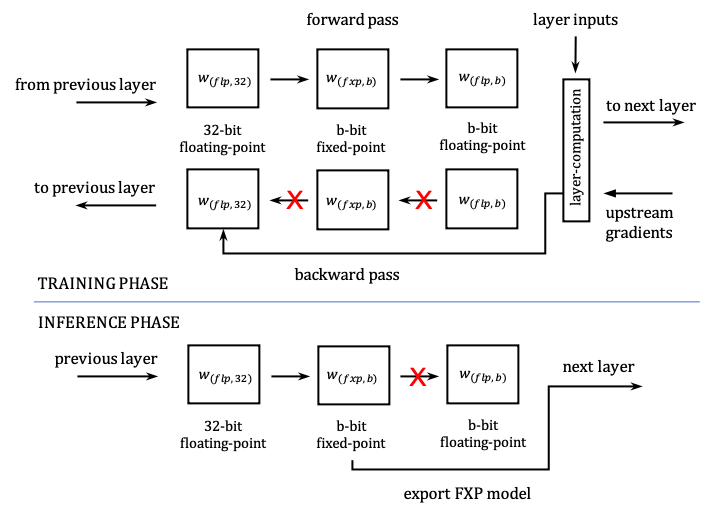}
  \vspace{-4mm}
  \caption{Proposed two-stage quantization approach to learn FXP weights in FLP training framework.}
  \vspace{-4mm}
  \label{figure:qat}
\end{figure}
\vspace{-2mm}
\subsection{Weight quantization}
\textbf{Using squashed weight distribution (SQWD)}, weights can be re-parametrized to a finite range (\textit{e.g.} [-1,1]). This can be achieved by means of a non-linear function (\textit{i.e tanh}) that maps weight \textit{(w)} to \textit{($\hat{w}$)} in uniform distribution~\cite{strom2022squashed}. A regularization loss limits the distribution-spread to asymptotes of \textit{tanh} for efficient mapping. Subsequently, \textit{$\hat{w}$} is quantized to desired $b$-bit precision using eq.~\ref{eq:float-quant}, where $w_{(flp,b)}$ refers to FLP weight \textit{w} with \textit{b-}bit precision. 
\begin{equation} 
\label{eq:float-quant}
  w_{(flp,b)} = (\text{round}[2^{b-1}(\hat{w} + 1) - 0.5] + 0.5) * 2^{b-1} - 1 
\end{equation}
\textbf{Using absolute cosine regularization (ACR)}, QAT can be defined as a regularization problem. Most QAT techniques do not include explicit quantization loss in back-propagation~\cite{Nguyen2020QuantizationAT}. Conversely with ACR, the model parameters are weighted using an absolute-cosine function that represents quantization loss. The cosine function has $2^b$ zero-points for \textit{b}-bit quantization. We constrain our weights to the same finite range as SQWD to establish a fair comparison.
\begin{figure}[tb]
  \centering
  \includegraphics[trim={0mm 0 0mm 0},clip,width=0.95\columnwidth]{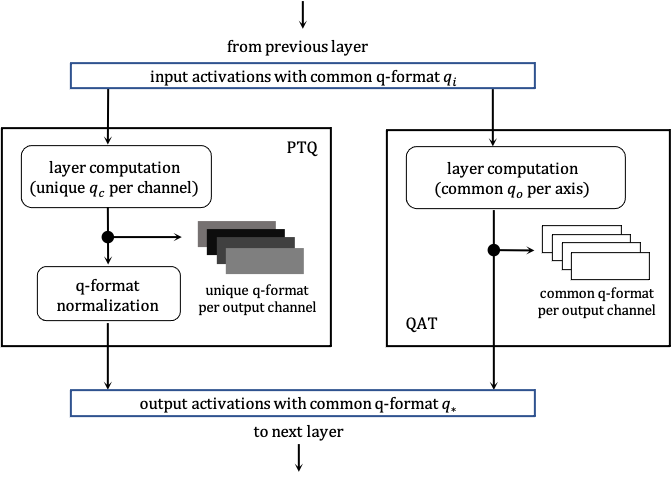}
  \vspace{-4mm}
  \caption{Activation normalization between two consecutive layers with PTQ. QAT limits the use of different q-formats eliminating normalization.}
  \vspace{-4mm}
  \label{figure:q-format}
\end{figure}
\vspace{-2mm}
\begin{align}
  w_{(fxp, b)} &= \text{round}[2^{b-1} * (w_{(flp, 32)} + 1.0)] \label{eq:quantize_fn} \\
  w_{(flp, b)} &= w_{(fxp, b)} * 2^{b-1} - 1.0 \label{eq:dequantize_fn}
\end{align}
\textbf{Proposed approach with FXP} Either in SQWD or ACR, a FLP quantizer is used to quantize a weight \textit{(w)} from full-precision FLP \textit{$(w_{(flp, 32)})$} to $b$-bit FLP \textit{$(w_{(flp, b)})$}. For a FXP inference with no quantization error, a trivial solution would be to replace the FLP quantizer in eq.~\ref{eq:float-quant} with a FXP quantizer. However, FXP arithmetic has limited support on GPUs which makes training large ML models challenging. To combat this, we propose a two-stage quantizer (shown in figure~\ref{figure:qat}) that maps \textit{w} from full-precision FLP $(w_{(flp, 32)})$ to \textit{b-}bit FXP $(w_{(fxp, b)})$ using eq.~\ref{eq:quantize_fn} and to $b$-bit FLP $(w_{(flp, b)})$ using eq.~\ref{eq:dequantize_fn}. This facilitates training in FLP arithmetic, albeit using FXP quantization. During back-propagation we use straight-through-estimators~\cite{bengio2013estimating} for gradient calculation, to overcome the non-differentiability of the round [.] operation. For inference, we use eq.~\ref{eq:quantize_fn} to export parameters to FXP. With a one-to-one mapping between quantized FLP and FXP representations, we can eliminate errors from PTQ.
\vspace{-3mm}
\subsection{Input and activation quantization}
\label{sec:input-quantization}
\vspace{-2mm}
Inputs to various model components form the remainder of inference arithmetic. In the context of hidden layer activations, deep models have a cascaded bit-growth problem due to the large number of multiply and accumulate (MAC) operations. To mitigate this, we placed fake quantization nodes between consecutive layers to scale down precision, and used a clipped ReLu with linear scaling to squash activations to smaller $[-1, 1]$ range, prior to quantization with eq.~\ref{eq:quantize_fn} and eq.~\ref{eq:dequantize_fn}. On the other hand, to accommodate wider numerical range of feature inputs, we traded-off bit precision on fractional part. The number of bits used to represent fractional part is termed as \textit{q-format}. A slightly modified quantization approach with \textit{q-format} for a feature \textit{f} is proposed in eq.~\ref{eq:quant-inp-eqn} and \ref{eq:de-quant-inp-eqn}, where $c=$ 0.5 if $f<0$ else -0.5.
\begin{align}[h]
  f_{(fxp, b, q)} &= \min[\max(f * 2^{q} + c, -2^{b-1}), 2^{b-1} - 1] \label{eq:quant-inp-eqn} \\
  f_{(flp, b, q)} &= f_{(fxp, b, q)} * 2^{-q} \label{eq:de-quant-inp-eqn} 
\end{align}
\section{FIXED-POINT INFERENCE} 
\label{sec:fixed-point inference}
\vspace{-2mm}
This section describes how QAT improves FXP inference by eliminating a key trade-off typical of post-training-quantized models, and leverages increased parallelization with low-bit inference on SIMD architectures. Additionally, we describe and propose a solution for archetypal saturations  in low-bit accumulators. The discussion is tailored in the scope of ARM NEON chipsets, a popular choice for FXP inference.
\vspace{-3mm}
\subsection{q-format normalization}
\vspace{-1mm}
FXP inference setups - specifically designed for PTQ assign a different q-format \textit{($q_{c}$)} per axis for model parameters to reduce quantization-error, resulting in mixed representations within the model. The number of representations scale with number of layers, and incur additional memory to store q-format mappings. To prevent this, activations are normalized during runtime to a uniform \textit{$q_{*}$} prior to subsequent layer. These normalization operations roughly consume 10\% cycles in inference. With QAT, we impose training time constraints to learn model parameters with a common q-format \textit{($\hat{q}$)} thus eliminating additional cycles spent in normalization, speeding up inference. This difference between PTQ and QAT is illustrated in figure~\ref{figure:q-format}.
%
%
\vspace{-3mm}
\subsection{Increased parallelization with QAT on SIMD}
\begin{figure}[h]
  \centering
  \includegraphics[trim={0.5mm 0 1mm 0},clip,width=0.6\columnwidth]{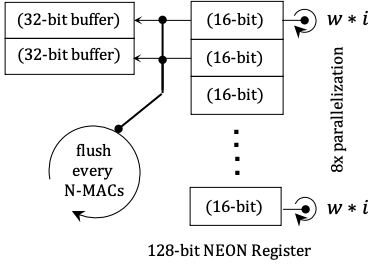}
  \vspace{-4mm}
  \caption{Proposed two-tier accumulator-buffer mechanism that prevent saturations in low-bit accumulator.}
  \vspace{-2mm}
  \label{figure:2TAB}
\end{figure}

\vspace{-1mm}
ARMv7a/v8a is the most adopted architecture for embedded devices using FXP arithmetic. The advanced-SIMD extension (NEON) in these architectures accelerates processing by parallelizing load, store, multiply, and accumulate operations. The \textit{vmlal\_s8} NEON intrinsic~\cite{neon-vmlal} operating on $128$-bit wide NEON registers supports up to eight operations in a single cycle. During inference, MACs make up to 90\% of computations, and we leverage parallelization capability of NEON to parallelize MACs and achieve faster inference. To fully utilize the $8\times$ parallelization, we limit the accumulator size in MAC operations to low $16$-bits, since $16\times$8=128, shown in figure~\ref{figure:2TAB}. The use of low-bit accumulator is enabled through FXP-QAT by training low-bit precision models. The $16$-bit accumulator is scaled-down prior to next layer to avoid bit-growth explosion. 
\vspace{-3mm}
\subsection{Saturations in low-bit accumulators}
\vspace{-1mm}
While smaller accumulators speed up inference, they saturate faster. The $16$-bit NEON accumulator can saturate in as little as 2 and 4 MACs using $8$-bit and $7$-bit models. Although we can employ QAT to train models up-to ultra-low-bit precision (\textit{i.e.} $4$-bit), model topologies influence saturations. Specifically, wide models involve greater MACs per activation and are more likely to saturate accumulators. Saturations are a significant problem for the model performance since they corrupt downstream model propogation. For the keyword-spotting problem we are interested in, we observed that 0.1\% saturations caused a $\sim$30\% degradation in keyword detection score. As a solution, we propose a two-tier accumulator-buffer (2T-AB) mechanism that mitigates this problem while maintaining benefits from 8$\times$ SIMD parallelism. As opposed to performing all the required MACs for an activation using a single accumulator, we propose to periodically flush the $16$-bit accumulator to a wider $32$-bit fixed-width buffer. In table~\ref{table:saturations}, we profile $6$-bit KWS model and show 2T-AB reduces the number of saturations. We observe that sparsely flushing (\textit{i.e} every 256 MACs) reduces saturations by 89\% while flushing every 64 MACs completely eliminates saturations.
\begin{table}[tb]
\vspace{-1mm}
\caption{Number of activations corrupted by saturations at different convolutional layers with and without 2T-AB.}
\vspace{1mm}
\renewcommand{\arraystretch}{1.25} 
\resizebox{\columnwidth}{!}{
\label{table:saturations}
\begin{tabular}{ccccccc}
\hline
\multirow{2}{*}{kernel size} &
  \multirow{2}{*}{\begin{tabular}[c]{@{}c@{}}\# MACs \\ per activation\end{tabular}} &
  \multirow{2}{*}{\# activations} &
  \multicolumn{4}{c}{\begin{tabular}[c]{@{}c@{}}\# activations corrupted by saturations \\ at flushing cadence\end{tabular}} \\ \cline{4-7} 
            &        &       & None & 256 MACs & 128 MACs & 64 MACs \\ \hline
(3, 4, 32)  & 384    & 10,880   & 2,598 & 237      & 47      & 0       \\
(4, 4, 32)  & 512    & 2,240   & 2,090 & 276      & 27      & 0       \\
(7, 4, 40)  & 1,120  & 128    & 128  & 12        & 0       & 0       \\
(1, 1, 128) & 128    & 160   & 39    & 6        & 0       & 0       \\
(1, 1, 160) & 160      & 3   & 3    & 3        & 2       & 0       \\ \hdashline
 TOTAL           &        &       & 4,858 & 534      & 76      & 0       \\ \hline
\end{tabular}}
\vspace{-5mm}
\end{table}
\vspace{-3mm}
\section{EXPERIMENTAL STUDY}
\label{sec:exp-setup}
\vspace{-2mm}
We study the proposed fixed-point QAT technique using a KWS model for resource constrained devices. We evaluate our approach using quantized models trained with our approach against an unquantized full-precision model in terms of \textit{i)} accuracy \textit{ii)} latency. 
\vspace{-3mm}
\subsection{Datasets}
\label{ssec:datasets}
\vspace{-1mm}
We conducted our experimentation using an in-house KWS dataset ($D_1$) and the Google speech commands v2 dataset ($D_2$)~\cite{warden2018speech}. $D_1$ consisted of 12.5K hours of de-identified labeled audio from far-field and near-field mobile phones and was split into training (10K hours) and testing (2K hours) while $D_2$ comprised of 105K one-second audio snippets sampled at 16kHz, containing 35 different keyword utterances recorded in natural environments~\cite{warden2018speech}. We used the official train/validation/test splits for $D_2$.
\\
The audio in $D_1$ and $D_2$ was down-sampled to 16kHz, converted to single channel, and segmented using a 25ms analysis window with a 10ms shift. While audio files have varying length, we used a segment length of 76 to generate fixed length inputs. Subsequently, we extracted and concatenated 64-dimensional log-mel filter bank energies (LFBE-64) to form 2-D feature inputs to our convolutional KWS model. 
\vspace{-3mm}
\subsection{Model configurations}
\label{ssec:model-config}
\vspace{-1mm}
We use a 2-D fully convolutional model for all experiments. The model contains five trainable consecutive convolutional blocks preceded by a non-trainable normalization layer. Convolutional blocks vary across the network in terms of kernel size, yet each convolutional layer is followed by batch normalization and ReLu nonlinearity as activation. The output is a softmax classification layer representing the posterior probability of keyword detection. The model has a total of 199K learnable parameters and  uses 125K flops.
\begin{table}[htb]
\caption{Relative false discovery rate (FDR) of different QAT models on dataset $D_1$ and their relative execution times (s) as measured on ARM CortexA53.} 
\vspace{1mm}
\centering 
\renewcommand{\arraystretch}{1.3} 
\resizebox{\columnwidth}{!}{
\label{table:overall-results}
\begin{tabular}{ccccc}
\hline
model         & bit-width & inference-format & relative FDR & relative exec-time \\ \hline
baseline      & 32-bit    & FLP              & 1.0          & 1.0                    \\
optimized-PTQ & 16-bit    & FXP              & 1.109        & 0.45                   \\
SQWD (QAT)    & 8-bit     & FXP              & 0.944        & 0.312                  \\
ACR (QAT)     & 8-bit     & FXP              & 0.960        & 0.312                  \\ \hline
\end{tabular}}
\vspace{-4mm}
\end{table}
\vspace{-3mm}
\subsection{Training configurations}
\label{ssec:training}
We standardize data prior to training using statistics computed from the training set and retain them for evaluation. We train all models in TensorFlow (2.4) for 1M steps and use Radam optimizer~\cite{Liu2020OnTV} to update model parameters. We increase learning rate to 1e-3 for the first 10$\%$ of total steps, and decay it linearly to 1e-5 thereafter. We use cross-entropy loss as our primary training objective, along with additional loss terms specific to different QAT paradigms. We use a batch size of 1500, and distribute training across 8 V100 GPUs. 
\vspace{-3mm}
\subsection{Evaluation metrics}
\label{ssec:eval-metrics}
We measure model accuracy using end-to-end Detection Error Tradeoff (DET) curves, which describe models' False Rejection Rate (FRR) vs. False Discovery Rate (FDR). Similar to~\cite{Zeng2022Sub8Q} we normalize the axes of DET curves and report relative FDR to full-precision floating-point models' FDR when relative FRR is set to 1.0. We measure inference speed using a single core CortexA53 based MT8163 processor specifically compiled for ARMv7a. We use an audio stream of length 90\textit{s} containing multiple instances of keyword, and run it continually on the processor against our KWS model to measure execution time. Similar to FDR, we normalize and report relative execution times to floating point model running on non-SIMD architectures. 
\begin{table}[h]
\vspace{-3mm}
\caption{Relative execution time (s) of different inference models on ARM CortexA53 versus flushing cadences to study 2T-AB.}
\vspace{1.5mm}
\label{table:cpu-exec-time}
\renewcommand{\arraystretch}{1.1} 
\resizebox{\columnwidth}{!}{
\begin{tabular}{ccccc}
\hline
model &
  \begin{tabular}[c]{@{}c@{}}inference-setup\\ (format, architecture)\end{tabular} &
  \begin{tabular}[c]{@{}c@{}}flushing \\ cadence\end{tabular} &
  \begin{tabular}[c]{@{}c@{}}degree of\\ parallelization\end{tabular} &
  \begin{tabular}[c]{@{}c@{}}relative\\ exec-time (\%change)\end{tabular} \\ \hline
baseline                   & FLP; non-SIMD              & None     & 1x                  & 1.0 (NA)           \\
optimized-PTQ              & FXP; SIMD                  & None     & 4x                  & 0.450 (55\%)    \\ \hdashline
\multirow{7}{*}{6-bit QAT (SQWD)} & \multirow{7}{*}{FXP; SIMD} & None     & \multirow{7}{*}{8x} & 0.312 (68.8\%) \\
                           &                            & 512 MACs &                     & 0.309 (69.1\%) \\
                           &                            & 256 MACs &                     & 0.309 (69.1\%) \\
                           &                            & 128 MACs &                     & 0.312 (68.8\%) \\
                           &                            & 64 MACs  &                     & 0.356 (64.4\%) \\
                           &                            & 32 MACs  &                     & 0.363 (63.7\%) \\
                           &                            & 16 MACs  &                     & 0.407 (59.3\%) \\ \hline
\end{tabular}}
\vspace{-6mm}
\end{table}
\subsection{Results and discussion}
\vspace{-1mm}
We train our KWS model at different precision using proposed QAT approach and refrain from changing model topology to compensate for performance at low-precision. We first trained a model in full-precision FLP \textit{(baseline)}, and performed post-training-quantization that includes common PTQ optimizations not limited to dynamic-q-factor selection, run-time normalization to reduce quantization losses \textit{(optimized-PTQ)}. We also trained two models using the two proposed FXP-QAT techniques (\textit{i.e} SQWD, ACR) at 8-bit precision. To run FXP inference, we export the model parameters to FXP with eq~\ref{eq:quantize_fn} and use a MT8163 processor compiled for ARMv7a. In table~\ref{table:overall-results}, we report relative FDR numbers for the above 4 models on dataset $D_1$. We observe that despite the optimizations, the PTQ models' performance deteriorates compared to the baseline, while the models trained with fixed-point QAT perform better than the baseline at half the precision. We attribute the performance improvement to additional regularization introduced by QAT. We also report the execution times of these four models (FLP using \textit{x86} and FXP using \textit{ARM CortexA53}) and observe that $16$-bit FXP-PTQ models, and FXP-QAT models are 55\% and 68\% faster compared to FLP baseline. 
Previously in table~\ref{table:saturations} we showed how proposed 2T-AB mechanism reduces saturations, but we note that the occasional flushing adds additional instructions during inference; although, the overhead is minimal compared to 8$\times$ speed up that we are able to realize. From table~\ref{table:cpu-exec-time} we see that a sparse frequency of 512, 256 or 128 MACs there is no observable slowdown, but a dense flushing at 16 MACs increases by a relative 10\%. Using dataset $D_2$ we present some ablation studies with in table~\ref{table:google-speech} and show that we are able to achieve comparable performance to floating-point accuracy at low-bit ($6$-bit) and ultra-low-bit ($4$-bit) precision with either SQWD or ACR. We also report that we could trade off precision on model weights than activations increased sensitivity towards activations. We confirm through our that we are able to successfully train FXP-QAT models that are faster and accurate than FLP counterparts.
\vspace{-2mm}
\begin{table}[tb]
\caption{Model accuracy on dataset $D_2$ at different weight (w) and activation (a) precision (in bits). Table a) employs QAT using ACR, and table b) uses SQWD. The full-precision floating point accuracy of the model on this dataset is 90.7\%.}
\label{table:google-speech}
\vspace{2mm}
\resizebox{\columnwidth}{!}{
\renewcommand{\arraystretch}{1.1} 
\begin{tabular}{ccccccccccccc}
\cline{1-6} \cline{8-13}
\multicolumn{1}{c|}{\diagbox[width=10mm, height=8mm,dir=NW]{w}{a}} & 8    & 7    & 6    & 5    & 4    & & \multicolumn{1}{c|}{\diagbox[width=10mm, height=8mm,dir=NW]{w}{a}} & 8    & 7    & 6    & 5    & 4    \\ \cline{1-6} \cline{8-13} 
\multicolumn{1}{c|}{8}   & 92.8 & 92.8 & 92.3 & 91.4 & 91.4 & & \multicolumn{1}{c|}{8}   & 91.6 & 91.4 & 91.4 & 91.1 & 90.4 \\
\multicolumn{1}{c|}{7}   & 92.4 & 92.9 & 92.7 & 91.9 & 91.6 & & \multicolumn{1}{c|}{7}   & 92.0 & 91.2 & 91.5 & 91.1 & 90.9 \\
\multicolumn{1}{c|}{6}   & 92.5 & 92.2 & 92.7 & 92.1 & 91.2 & & \multicolumn{1}{c|}{6}   & 91.3 & 91.6 & 91.7 & 91.3 & 91.3 \\
\multicolumn{1}{c|}{5}   & 92.6 & 92.5 & 92.8 & 92.3 & 90.9 & & \multicolumn{1}{c|}{5}   & 91.3 & 91.5 & 92.3 & 91.3 & 91.9 \\
\multicolumn{1}{c|}{4}   & 93.1 & 92.7 & 92.5 & 92.3 & 91.8 & & \multicolumn{1}{c|}{4}   & 91.5 & 91.7 & 91.6 & 91.6 & 91.3 \\ \cline{1-6} \cline{8-13} 
\multicolumn{13}{c}{} \\[-1ex]
\multicolumn{6}{c}{(a)}                                     & & \multicolumn{6}{c}{(b)}
\end{tabular}
}
\vspace{-6mm}
\end{table}
\vspace{-3mm}
\section{CONCLUSIONS AND FUTURE WORK}
\label{sec:concl}
\vspace{-1mm}
In this paper, we presented a method for training and developing FXP models that uses floating-point arithmetic during training, but has a one-one mapping and eliminates conversion errors post-training. We applied this method in conjunction with two different QAT techniques to learn low-precision model parameters and also propose methods that apply QAT over transient variables. We test our approach in the scope of keyword-spotting that often require accurate and low-latency inference. We initially demonstrate and show that our 8-bit FXP QAT models outperform 32-bit FLP counterparts. Encouraged by the promising results, we further reduced precision, and observed that our method has a strong generalization capability allowing us to train models up to ultra-low-bit precision ($4$-bit) with minimal accuracy degradation. Besides accurate, our model is more efficient than traditional models, achieving ~68\% and 30\% speed up in execution time compared to floating-point inference, or the traditional combination of post training weight quantization, followed by fixed-point inference. In conclusion, we argue that it is possible to reduce on-device computation processing time, without sacrificing accuracy, which is especially important for always-on keyword spotting systems. In the next iteration of our work we plan to explore how our approach scales to large models and different problem spaces like computer vision and complex tasks like speech recognition. Additionally, we plan to combined our work with other compression techniques like pruning, distillation, and neural architecture search to generate efficient models. We believe that our findings will broadly impact the community focused on deep learning on resource constrained devices.
\vfill
\bibliographystyle{IEEEbib}
\bibliography{refs}
\end{document}